\documentstyle[12pt]{article}
\begin{document}
\setlength{\baselineskip}{0.30in}

\newcommand{\beq}{\begin{equation}}
\newcommand{\eeq}{\end{equation}}

\newcommand{\bi}{\bibitem}

{\hbox to\hsize{September 1995 \hfill FTUV/95-02; IFIC/95-44}}

\begin{center}
\vglue .06in
{\Large \bf {Problems of Dark Matter}}\\[.5in]

{\bf A.D. Dolgov \footnote{Permanent address: ITEP, 113259, Moscow,
Russia.} }
\\[.05in]
{\it{Instituto de Fisica Corpuscular - C.S.I.C.\\
Departament de Fisica Teorica, Univesitat de Valencia \\
46100 BUrjassot, Valencia, Spain }}\\[.15in]

{Abstract}\\[-.1in]
\end{center}
\begin{quotation}
The data indicating existence of different forms of dark matter in the
universe as well as the role of this matter in structure formation
are briefly reviewed. It is argued that vacuum energy gives a dominant
contribution into the total energy density of the universe. The model
of structure formation with unstable tau-neutrino with MeV-mass and
KeV-majoron is described.
\end{quotation}

\newpage

It is commonly agreed now that the matter (or energy) in the universe
is predominantly (90\% or even more) invisible and that most
probably this invisible matter is not the normal baryonic staff that
we are familiar with. The agreement however does not go beyond this
point and as for the nature of this dark matter the problem remains
open and will probably be open for quite some time. There are some
hints from cosmology regarding the properties of dark matter. They are
given by direct astronomical observations and by the theory of large
scale structure formation in the universe. From the other side the
elementary particle theory gives plenty predictions about possible
new stable particles which might be constituents of dark matter.
Unfortunately these predictions do not directly follow from the well
established minimal $SU(3)\times SU(2)\times U(1)$ - theory but from
one or other higher energy extensions of this theory which have not
yet been verified by experiment. So at the moment we are unable to
chose different possibilities and have to rely only
on aesthetic properties of the model which may be grossly misleading.

The fraction of a particular form of matter in the universe is given
in terms of the dimensionless parameter
\beq{
\Omega_j = \rho_j / \rho_c
}\label{omega}
\eeq
where $\rho_j$ is the mass (energy) density of this form of matter and
$\rho_c$ is the critical or closure density,
\beq{
\rho_c = { 3H^2 m_{Pl}^2 \over 8\pi } = 1.88 h_{100}^2 \times
10^{-29} \,g/cm^3 = 10.5 h_{100}^2 \,KeV/cm^3
}\label{rhoc}
\eeq
Here $m_{Pl} = 1.22\times 10^{19}$GeV is the Planck mass (the
Newtonian gravitational constant is $G_N = m_{Pl}^{-2}$). The Hubble
constant $H$ is parametrized in terms of dimensionless quantity $h$:
\beq{
H = 100 h\,km/sec/Mpc
}\label{H}
\eeq
It is usually assumed that $ 0.4 < h < 1 $, but the recent
data~\cite{hubble} favor the values in the upper part of this interval,
$h=0.75-0.85$. The larger values of $h$ were rejected by many
astronomers because of the universe age problem (see below) but now
more and more evidence is accumulated in favor of that and the
attitude seems to be gradually changing.

If the Hubble constant and $\Omega_j$ are known, the universe age
can be expressed through them as
\beq{
t_u = {1 \over H} \int_0^1 {dx \over \sqrt{ 1 - \Omega_{tot} +
\Omega_{m} x^{-1} +  \Omega_{vac}x^2 }}
}\label{tu}
\eeq
where $\Omega_{m}$ and $\Omega_{vac}$
correspond respectively to the energy density of nonrelativistic
matter and to the vacuum energy density
(or, what is the same, to the cosmological constant);
$\Omega_{tot} = \Omega_{m} + \Omega_{vac}$,
and $1/H = 9.8 h_{100}^{-1}\times 10^9$yr.

On the other hand the universe age can be found from nuclear
chronology which uses measurements of the ratios of the long-lived
isotopes, $^{187}Re / ^{232}Th$ or $ ^{238}U / ^{235}U$, and from
the estimated ages of old stellar clusters. Both methods give close
results in the range (12 - 20) Gyr. The recent analysis gives
$t_u \approx 18$ Gyr, though somewhat smaller values are possibly not
excluded (for the review and the list of references see
e.g. paper~\cite{vdb}).

If the universe is dominated by nonrelativistic matter, as usually
assumed, the age is approximately equal to
$t_u \approx 9.8 h^{-1} / (1 + \sqrt \Omega_m/2 )$ Gyr. Even with
$\Omega = 0$ the large age, $t_U = 18\,Gyr$, and $h>0.75$ are inconsistent.
The observed tendency to large values of $H$ and
$t_u$ presents a very strong argument in favor of nonvanishing vacuum
energy. For $\Omega_{tot} = \Omega_{m} + \Omega_{vac} = 1$ eq.(\ref{tu})
gives
\beq{
t_u = {2 \over 3H \sqrt \Omega_{vac}} \ln { 1 + \sqrt \Omega_{vac} \over
  \sqrt { 1 -  \Omega_{vac} }}
}\label{tuvac}
\eeq
With $h = 0.75$ and $\Omega_{vac} = 0.8$ we get $t_u = 14$ Gyr.
If this is the case most of invisible energy in the universe
is just the energy of empty space, of vacuum. Unfortunately
our understanding of vacuum energy is very poor. Any reasonable
estimate of it gives the result which is some 50-100 orders of magnitude
higher than the observational limit $\rho_{vac} < 10^{-47}\,GeV^4$
(for the review see refs.~\cite{sw2,ad2}).
For example there are contributions from quark
and gluon condensates which are well established in QCD and which are
about $10^{-4}\, GeV^4$.
So there must exist a contribution into vacuum energy from
something not related
to quarks and gluons but with exactly the same magnitude and
the opposite sign. It is hard to
imagine an accidental cancellation with such an accuracy but no
dynamical mechanism has yet been found.  To
my mind the best possibility for solving the cosmological
constant problem is the adjustment mechanism~\cite{ad3,ad4}
(for recent papers and some references see~\cite{fujii}
and reviews~\cite{sw2,ad2}), which
ensures a cancellation of vacuum energy by an action of a new
field coupled to gravity. This cancellation is
generically not complete and a noncompensated amount of
$\rho_{vac}$ is always of the order of $\rho_c(t)$. In such
models vacuum energy and the energy of the new field are
essential at any stage of the universe evolution (in contrast to
the models with normal time independent vacuum energy) and this
might have an impact on primordial nucleosynthesis, structure formation,
etc.

Astronomical data on the amount of different forms of matter in the
universe can be summarized as following:
\begin{enumerate}
\item{}
The directly seen luminous matter contributes very little
to the total mass of the universe: $\Omega_{lum} =(3.5\pm 1)\times
10^{-3} h^{-1}$.
\item{}
Total amount of baryons found from primordial nucleosynthesis
is $\Omega_{bar}^{(NS)} = 4\times 10^{-3} \eta_{10} h^{-2}$ where
$\eta_{10} = 10^{10} (N_B / N_\gamma)$ with $\eta_{10} =  1-7$.
The recent analysis of the problem can be found e.g.
in refs.~\cite{ost,os,cst}.
At the moment there is no consensus whether $\eta$ is
relatively high, near the upper bound, or near the lower end of the
interval.
In the first case we should expect plenty of invisible baryons in the
universe while in the second all baryons may be visible. To resolve
this ambiguity it is very important in particular
to find the amount of primordial
deuterium. At the moment there are two conflicting sets of observations,
one gives by number
$D/H \approx 2\times 10^{-4}$~\cite{crw,sch} and requests a
small $\eta$, $\eta_{10} \approx 1$, while the other gives
$D/H = (1-2)\times 10^{-5}$~\cite{tf} and $\eta_{10} \approx 3$.
\item{}
Flat rotational curves observed in gravitationally
bound systems like gas around galaxies or galactic
satellites give, depending on scale, $\Omega_{rot}=0.1-0.3$.
Rotational velocities are measured for hundreds of galaxies
up to distances about
30 Kpc through HI hydrogen gas, up to 100 Kpc for hot X-ray gas
around galaxies and up to 200 Kpc for galactic
satellites~\cite{ash,cen,persic}. These flat rotational curves mean that the
gravitating mass is not confined inside the luminous region with the
size about 10 Kpc (for large galaxies) but linearly rises with the
distance, $M(r) \sim r$. No cutoff for this behavior has yet been
observed.
\item{}
The recent analysis~\cite{hdc} of peculiar velocities of about
3000 galaxies gives $\Omega_{lsf}=[b(0.74\pm 0.13)]^{5/3}$
where $b$ is the biasing parameter,
$b=(\delta\rho /\rho)_{vis}/(\delta\rho /\rho)_{tot} $. The value of
the latter is not known but we believe that it is close to one and
most probably $b\geq 1$.
\item{}
The lower limits on the universe age, $t_U > 12\,Gyr$, and on the
Hubble constant, $h > 0.7$, result in the bound
$\Omega_{matter} <0.2$, if cosmological constant is zero.
\item{}
Recent observations~\cite{wnef,bc} of hot X-ray gas in rich galactic
clusters showed a surprisingly large fraction of baryons with respect
to the total mass of the clusters. If the mass contained in the
clusters represents a
fair sample for the total mass in the universe then these data
together with the nucleosynthesis constraint on $\Omega_B$ put the
strong upper bound on the energy density in the universe:
$\Omega_{clustered} \leq 0.15 h_{100}^{-1/2} /(1+0.55h_{100}^{3/2})$.
This bound is evidently valid for the clustered matter and is not
applicable for the uniformly distributed one.
These data are probably a good indication on nonzero vacuum energy
$\Omega_{vac} \approx 0.8$.
It is uniformly distributed and for the typical cluster size about
1 Mpc does not contribute much inside this radius.
\item{}
Inflationary universe model predicts $\Omega = 1 \pm 10^{-4}$. Recently
there appeared attempts~\cite{bgt} to reconcile inflation with
$\Omega \neq 1$, which were stimulated by the new astronomical data.
These attempts however do not look as natural as the original
inflationary prediction since they request a tuning of inflationary
and postinflationary stages in such a way that $\exp(2H_I\tau) =
(T_R/T_0)^2 (\Omega_0^{-1} -1)^{-1}$, where $T_R$ is the reheating
temperature after inflation, $T_0= 2.7\,K$ is the present-day
temperature of the cosmic microwave radiation, and $\tau$ is the
duration of inflation. This condition looks very strange
because physically these epochs are not related. To my mind
$\Omega = 1$ remains a very strong prediction of inflationary
cosmology.

\end{enumerate}
At the present stage one cannot definitely claim what is the correct
value of $\Omega$. I think that a nonzero value of $\rho_{vac}$ should
be seriously considered. To my mind the best choice is the following:
$\Omega_{bar} \approx \Omega_{vis} \approx (0.3-0.5)\%$;
$\Omega_{DM} = 0.1-0.2$, and $\Omega_{vac} = 0.8-0.9$ with
$\Omega_{tot} = 1$.

Even if one believes that vacuum energy contributes (80-90)\% to
the total energy density in the universe, still there should be some
other unknown form of invisible matter which provides flat rotational
curves, large scale flows and contributes the remaining (10-20)\% into
$\Omega_{tot}$. Nucleosynthesis constraint does not permit this matter
to be baryonic and we do not know what it is. The only hints about
its nature come from elementary particle theory and from the theory of
large scale structure formation. Minimal supersymmetric extensions of the
standard model predict existence of a massive stable particle with
$m=O(100\,GeV)$. The calculated cosmic abundance of these particles
naturally give $\Omega_{LSP} $ close to unity (here LSP stands for
lightest supersymmetric particle). This coincidence is
quite impressive because apriori one could expect  $\Omega_{LSP} $
different from 1 by many orders of magnitude. This makes LSP a very
good candidate for dark matter particle (for a recent review see~\cite{kane}).

Still one may express some doubts regarding this possibility.
Stability of LSP is ensured by R-parity which is conserved in simple
supersymmetric models. However we know from the history
of particle physics during last half of century that
the only conservation laws survived which were protected by a well
justified theoretical principle, like e.g. CPT-theorem or electric
charge conservation (protected by gauge invariance in QCD). As for
R-parity no such principle naturally appeared and thus one should
expect that it is nonconserved. In that case LSP's are unstable and
if their life-time is cosmologically small, we have to look for other
possible candidates for dark matter particles. These could be axions,
massive neutrinos, or maybe some other more exotic forms of matter.
A classical field which probably compensates nonzero vacuum energy
is a very interesting possibility. Models with broken R-parity
which may kill stable LSP's, simultaneously give birth to another possible
candidate for dark matter. R-parity breaking is realized through
spontaneous breaking of leptonic charge conservation~\cite{jv1,cmp}
associated with global symmetry group $U(1)_L$.
Spontaneous breaking of a global symmetry group
gives rise to appearance of a (pseudo)goldstone boson, majoron.
If there is also an explicit symmetry breaking, the majorons would
become massive and, if cosmologically long-lived, could be
constituents of dark matter~\cite{bv,dpv}.

Some of dark matter candidates, like e.g. massive stable neutrinos,
may be rejected on the basis of the
theory of large scale structure formation. However
not all the basic assumptions of this theory are solid enough to make
really strong conclusions. While practically everybody agrees that
the universe structure has been formed
as a result of evolution of initially small fluctuations under the
action of gravitational forces, there is no consensus about the shape of
the spectrum of initial density perturbations. Introducing dimensionless
quantity $\delta = \delta \rho /\rho$ we can present the power spectrum
in the form
\beq{
\langle \delta^2 (x) \rangle = \int dk f^2(k)
}\label{delta2}
\eeq
(assuming that Fourier amplitudes are delta-correlated; one more
assumption which may be questioned). Usually for $f(k)$,
the most simple form is
taken namely it is assumed that it does not contain any
dimensional parameter, $f(k)^2 \sim 1/k$. This corresponds to flat or
scale-free spectrum proposed by Harrison~\cite{har} and
Zeldovich~\cite{ybz2}. Inflationary models gave theoretical
justification to this form of spectrum and
now it is commonly used as the basic spectrum
in calculations of structure formation. Sometimes as a simple
generalization an  arbitrary power law spectrum is considered,
$f^2 \sim 1/k^n$. The spectrum with
$n\neq 1$ may appear in some inflationary models~\cite{kf} but with $n$
not much different from unity. If we permit in principle an existence
of primordial spectrum with an arbitrary $n$, introducing in this
way a new scale to the theory, a combination of several
terms with different $n$ as well as a more complicated functions
are also permitted but without a guiding principle for choosing a
particular form of the spectrum, the theory would completely loose
its predictive power. So we have to keep in mind that
the judgement on the possible dark matter
particles is made under assumption that the spectrum of perturbations
has simple scale-free form.

The simplest model which was successfully used for description of
structure formation until recently was the one with a single component
cold dark matter particles (like LSP or axion)
and flat spectrum of initial perturbations (for the review see e.g.
refs.~\cite{mixed,cdm}). The model
reasonably well described galaxy distribution at the scale of tens
megaparsec with the only free parameter, the overall normalization
of the spectrum. However the COBE measurements~\cite{cobe} has
fixed the normalization at large scale end of the spectrum and with
this normalization the predictions at smaller scales became approximately
twice above the observations. An evident possible cure is to change the
shape of the spectrum of primordial fluctuations. It seems to be
a very interesting possibility but since this idea is not
supported by the inflationary scenario, the major line of investigation
goes along consideration of different forms of dark matter
with the same flat spectrum of density perturbations or maybe an
introduction of the cosmological constant.
The basic idea of all these models was to
suppress the power of the evolved spectrum at smaller scales relative
to that at larger (COBE) scales.

One possibility is to shift the epoch of matter dominance (MD)
to a later stage in a simple cold dark matter model.
Since the characteristic scale at which perturbations started to rise
in this model is determined by the horizon size at the onset of the
MD stage, shifting it to a later moment gives less time for rising of
the fluctuations and correspondingly less power at galactic
and cluster scales. This goal can be achieved if one assumed
that universe is open so that $h_{100}^2 \Omega \approx 0.2 $.
However the low value of $\Omega$
is disfavoured by inflationary scenarios which (at least in simple
versions) predict $\Omega = 1$. One can recover this prediction
of inflation in the universe with low matter density if the
cosmological constant $\Lambda$ (or in other words, vacuum energy) is
nonzero. As we have mentioned above
the recent data indicating a rather high value of the Hubble constant,
$h_{100} \approx 0.8$, support the idea of nonzero $\Lambda$
with the fraction of the vacuum energy $\Omega_{vac} = 0.8$.
The models with nonzero
$\Lambda$ give a satisfactory description of the observed
structure~\cite{vacstruc} with flat spectrum and COBE normalization.

A mixed (hot+cold) dark matter scenario can
also do the necessary job of diminishing the power at small
scales because
(initially flat) perturbations in hot dark matter are erased at
scales smaller than $10^{14} M_{\odot}$ by free streaming if the
dark matter is collisionless as is the case of neutrinos. A good
description of the structure requests 70\% of CDM and
30\% of HDM~\cite{mixed}. It
gives even better description if there are two equal mass neutrino
species each with the mass 2.5 eV~\cite{prim} as is suggested by
the recent indications of neutrino oscillations by the Los Alamos group.

Recently there appeared a renewed interest to the idea of
structure formation with unstable particles~\cite{unstable}.
It is assumed in these
models that there exists a massive long-lived particle, usually
tau-neutrino with the mass in MeV range which decayed into massless
species at the epoch
when the mass density of the parent particles dominated the
energy density of the universe. Correspondingly the present-day
energy density of relativistic particles would be bigger than in the
standard scenario and the onset of MD stage would take place later.

The common shortcoming of these models is that they all demand a
certain amount of fine-tuning. Generally one would not expect that
the contribution from hot and cold dark matter into the universe
mass are about the same, they may differ by many
orders of magnitude.
One would also suspect that the vacuum energy which remains
constant in the course of the universe expansion is by no means related
to the critical energy now, which goes down with time as
$m_{Pl}^2 /t^2$. However adjustment mechanism \cite{ad3,ad4}
may possibly solve this problem.
The models with unstable particles mentioned above are also based on
the assumption of two independent components with a close
contribution into $\Omega$: the massive unstable
particles themselves and unrelated cold dark matter. This is
definitely unnatural and this shortcoming stimulated search for
other models. Recently in ref.~\cite{wssd} a return to to the
universe with a single cold dark matter component (of course
except for baryons) was advocated. For successful description
of the observed structure the authors need the power index of the
spectrum $n=0.8-0.9$, a low value of the Hubble constant,
$h = 0.45-0.5$, and
a large contribution of tensor perturbations (gravitational waves)
into quadrupole fluctuation of background radiation temperature,
$C_T /C_S = 0.7$.

The old idea of mirror world \cite{mw} which is coupled to ours
(almost) only through gravity was recently revived \cite{bm,sfv}
for possible explanation of the puzzles in neutrino physics
and as a possible source of dark matter in the universe. The
model of ref.\cite{bm} predicts an existence of a heavy sterile
neutrino with mass in KeV range which might be a warm dark matter
particle and a light (but massive) usual neutrino with mass
about 10 eV for a hot dark matter particle. The relic abundances
of these two neutrinos are predicted to be close to each other.

Another attempt to overcome unnaturalness of multicomponent dark
matter model has been done in paper~\cite{dpv}.
In this paper a model is considered in which unstable particles and
the particles of cold (or possibly warm) dark matter are closely
connected. In fact the decay of the former produces particles of the
present-day dark matter. A necessary background model of this kind
in particle physics was proposed some time ago~\cite{jv1,cmp} as
an attempt to find a phenomenologically acceptable description of
R-parity breaking. The underlying mechanism is the
spontaneous breaking of leptonic charge conservation at
electroweak scale. The model contains a Majorana type
tau-neutrino with the mass around MeV which decays into massive but
light majoron $J$ with mass in KeV region. Life-time with respect to
this decay, as estimated in refs.~\cite{jv1}, could be of order of
years depending on the values of parameters, i.e. in the interval
interesting for structure formation. It is worth noting that there is no
stable SUSY particle in this model
so the dark matter cannot be associated with it but the model itself
produce a candidate for dark matter, namely a massive majoron.
The model possesses some features like
sufficiently large diagonal coupling of $\nu_\tau$ to majorons or
selfinteraction of majorons which  rather naturally permit
to resolve appearing cosmological problems in particular the
problem of extra massless particle species during primordial
nucleosynthesis. The cosmological properties
of this model are rather unusual and are interesting by themselves.
Dark matter particles (majorons) in this model are strongly
self-interacting and thus the structure formation in this model
is different from the traditional one with collisionless dark matter
particles~\cite{hall,mem,lss,dpv}. In particular the shape of galactic halo
would depend on the dark matter selfinteraction. The
recent data~\cite{shape} might give an indication that collisionless
dark matter gives a poor description of the shape of halo in
dwarf galaxies.

At first sight the model of ref.~\cite{dpv} encounters very serious
cosmological difficulties. Tau-neutrino with the mass in MeV range
and stable at nucleosynthesis time scale would strongly destort
primordial abundance of light elements~\cite{ckts,dr} and should
be excluded. However the Yukawa coupling of $\nu_\tau$ to majorons
helps to reduce the number density of tau-neutrinos down to a safe
value through the reaction $2\nu_\tau \rightarrow 2J$. In this model
the effective number of relativistic degrees of freedom during
nucleosynthesis may be even below 3 in agreement with the recent
claims~\cite{ost,hss}.

However a strong reduction of the number density of $\nu_\tau$ at
nucleosynthesis through the annihilation into majorons produces the
equilibrium amount of majorons such that the ratio of their number
density to that of photons becomes $n_J/n_\gamma =
0.5 (T_J /T_\gamma)^3$. By Gerstein-Zeldovich bound~\cite{gz,cml}
the mass of such particles cannot exceed roughly speaking 30 eV.
On the other hand for succesful structure formation we need
$m_J = O(KeV)$ and it looks as if the model does not work. Fortunately
this is not so because of relatively strong majoron
selfcoupling: $\lambda J^4$ with  $\lambda = 0.1-0.01$~\cite{bv}. This
interaction gives rise to majoron cannibalism through the process:
$4J\rightarrow 2J$. This process is fast enough to reduce the majoron
number density by an order of magnitude~\cite{dpv} and to permit the
majorons to have the mass appropriate for warm dark matter particles.

To conclude we definitely know that the world predominantly consists of
matter which is different from what we directly see around. We do not
know what kind of matter it is. Most probably this unknown matter
consists of several different components. Why these components have
similar energy densities in the present-day universe is also unknown.
One of the component is very probably just energy of vacuum.
The problem of vacuum energy is, to my mind, the central one in
cosmology and particle physics in particular in connection with dark
matter. A new massless field coupled only to gravity may solve the
cosmoogical constant problem and to contribute to the dark matter.
As for more traditional components which should be also present, even
if vacuum energy is nonvanishing, the search for low energy
supersymmetry is very important In particular it is essential to know
if R-parity is indeed conserved and the lightest supersymmetric
particle is stable. An improvement of accuracy in determination of
tau-neutrino mass may have an important impact on cosmology.
A related phenomenon is
the primordial nucleosynthesis because the produced abundance of light
element depend on $\nu_\tau$ mass if it is in MeV range. Hopefully in
the nearest future the ambiguity in determination of the amount of
primordial deuterium, which is very sensitive to the total density
of baryons will be setteld down. If one is optimistic one may
hope that the mistery of building blocks of the universe will be
resolved before the end of the next (or this, if one is superoptimistic)
century.

\section*{Acknowledgments}
This work was supported by DGICYT under grants
PB92-0084 and SAB94-0089 (A. D.).


\end{document}